# Measuring a moving target- Innovation studies in practice

## Abstract

This paper pays a tribute to Loet's work in a specific way. More than 20 years ago Loet Leydesdorff and myself designed a programme for future innovation studies "Measuring the knowledge base - a programme of innovation studies". Although, the funding programme we envisioned eventually did not materialise, the proposal text set out the main lines of our research collaboration over the coming decades. This paper revisits main statements of this programme and discusses their remaining validity in the light of more recent research. Core to the Leydesdorff/Scharnhorst text was a system-theoretical, evolutionary perspective on science dynamics, newly emerging structures and phenomena and addressed the question to which extent they could be meaningful studied using quantitative approaches. This paper looks into three 'cases' – all examples of newly emerging institutional structures and related practices in science. They are all located at the interface between research and research infrastructures. While discussing how the programmatic ideas written up at the beginning of the 2000s still informs measurement attempts in those three cases, the paper also touches upon questions of epistemological foundations of quantitative studies in general. The main conclusion is that combining measurement experiments with philosophical reflection remains important.[1]

## Introduction

In the last decades, academia has fundamentally changed, not mainly *in its essence* (the fact that we think), but rather *in its practices* (the way we organise our thinking). By taking an evolutionary perspective in the analysis of the science system, we observe both a growth of the science system as well as more differentiation. Differentiation here concerns both the cognitive level (in terms of the emergence of new specialties and sometimes new fields and disciplines) as well as the organizational level (namely, institutions and other forms to organize research such as (funded) projects). As it happens in the evolution of any complex system (and I consider the science system to be such), we also observe counteracting trends to growth and differentiation. An important phenomenon is the emergence of new intermediaries and new coordinating levels.[2] One example of such a phenomenon are research infrastructures. Examples of research infrastructural institutions are libraries, archives, but also computing centres. Their role has changed from being purely supportive (e.g., responding to new technologies needed to perform research activities), to increasingly becoming co-creators in the knowledge production process (Morselli 2023).

This paper reflects on the extent to which such new coordinating structures can be captured in quantitative studies. It does so, by using concrete examples: (a) a research data archive,

---

[1] This paper is based on a presentation given at the workshop "Philosophy of Science Meets Quantitative Studies of Science" organized by Eugenio Petrovich in Turin in May 2024.
[2] Here I refer particularly to early theories of self-organisation and complexity such as those described in the book of Maturana and Varela from 1984 (1987) *The Tree of Knowledge*.

(b) the DARIAH European Research Infrastructure; and (c) research data management practices in a research project.

Looking at 'measuring attempts' around these examples, I touch upon the fundamental questions of why, what and how aspects of knowledge production can and should be measured. To that purpose, I'm using the perspective of the 'Measuring the Knowledge Base' programme (Leydesdorff, Scharnhorst 2003) which will be introduced in the next section as analytical lens for the three cases mentioned above. The paper ends with conclusions which discuss to which extent the programmatic text of Loet Leydesdorff and this author is still an interesting piece of communication to relate to, also for current quantitative studies.

## Measuring the knowledge base

The text 'Measuring the knowledge base – a programme of innovation studies' (Leydesdorff, Scharnhorst 2003)[3] was written in response to a call of the *Berlin-Brandenburgische Akademie der Wissenschaften* (BBAW) in the early 2000s. This call invited ideas for research lines which might become the subject for further funding. This purpose determined the programmatic nature of the text. Although the submission of myself and Leydesdorff was eventually not chosen for further funding, it represents a kind of reference point for our shared view on innovation studies and our further collaboration departed from the ideas laid out in this text.

The aspiration of our programme was to define a framework which enables meaningful measurement of newly emerging phenomena in the science system. Its central question was: Can the new mode of knowledge production be measured? The text departs from three core statements: (1) knowledge production is at the source of innovation, (2) innovation itself is the base for wealth and welfare in highly industrialised western countries, and (3) one needs to understand the mechanism of innovation better in order to support a science policy which guarantees prosperity and welfare. As authors we also demarcated our methodological approach: as organized knowledge production is codified in communication, these communications leave traces (think here publications, patents, or whatever other documentation will become available), and can be studied by means of communication theory – combining Luhmann's system theoretical approach (Luhmann 1990) to science and society with Shannon's mathematical theory of communication (Shannon, Weaver (1949). The problems the programme identified are (1) to choose the 'right' unit of analysis and (2) to connect empirical studies with a theoretical base.

At the beginning of the 2000s, academic institutions became increasingly interwoven and new forms of institutions emerged. The phrase *Knowledge Production Mode 2* (Gibbons et al. 1994) had been coined and largely influenced the studies of science and societies in the decades following its publication. With *Mode 2* it was indicated that the way how and in which institutional setting knowledge was produced had changed. However, not only did the production of knowledge change, the role of knowledge production for innovation (technological and economic) in society also changed. We conclude in the text that national welfare can by no means be organized solely by looking at national systems of innovation. The national economies throughout Europe and the rest of the world had become too interdependent. The text of the proposal argues that the traditional political economy needs to be expanded into a knowledge-based economy.

---

[3] The text was finished 2002, and posted as preprint to the arXiv in 2003.

The aspiration of our programme was to define a framework which enables meaningful measurement of newly emerging phenomena in the science system. It was firmly rooted in a system-theoretical approach both authors shared as well as in the belief that the science system is a complex system – new ways to study it were needed and to do so in a comprehensive way. While our proposal was methodological, centered in communication theory, we pointed out that any research on this matter also needed to be informed by historical and philosophical thinking, by theories from economic and political sciences, and by complex system and network theory.

> "As innovations take place at interfaces, the competitive advantages in a knowledge-based economy can no longer be attributed to a single node in the network. The networks coordinate the sub-dynamics of (i) wealth production, (ii) organized novelty production, and (iii) private appropriation versus public control." [Leydesdorff, Scharnhorst 2003, p. iv]

Loet Leydesdorff at the time of writing this text was member of the famous 'Science Dynamics' group at the University of Amsterdam. He was a recognized scientometrician, member of the editorial board of the journal *Scientometrics* since 1987. In the 1990's, together with Henry Etzkowski, he wrote about the *Triple Helix Model of Innovation* in a series of publications, which would go on to become the most highly cited in his career and effectively opened up a whole new research thread (Etzkowski, Leydesdorff 1995; 2000). In 1993, Loet also organized (together with Peter van den Besselaar) a conference on Evolutionary Economics in Amsterdam to further foster bridges between natural sciences and innovation studies (Leydesdorff, Van den Besselaar 1994). The 'Epilogue chapter' by Leydesdorff in the proceedings of this conference already calls for new models of technological change, and reaches out to theories of chaos, self-organisation and what later became called complexity theory (Leydesdorff 1994). The 'Measuring the knowledge base' text builds on those thoughts, aspiring to look beyond institutions in the analysis and emphasizing the challenges that arrive when innovation is understood as a phenomenon born from networks of various nature.

If I remember correctly, Loet and I met at the first ISSI conference organized by Hildrun Kretschmer (Glänzel et al. 2022) in Berlin 1990, later again at the Science and Technology Indicator conference in Leiden 1991 (Bonitz et al 1992), and at the above mentioned 'Evolutionary Economics conference' in Amsterdam 1993 (Bruckner et al. 1994). I was a young researcher, who had just started an academic career in the German Democratic Republic (GDR) Academy Institute with the name "Institute for Theory, History and Organisation of Science" (*Institut für Theorie, Geschichte und Organisation der Wissenschaft (ITW)* (Kröber 1998; Laitko 2007; 2018; Cain 2021) when the *Wende*[4] catapulted everybody from the socialist reality of the East into the capitalist market economies of the West. This transition opened new possibilities but, for the GDR academia, also extinguished almost all Academy institutes and expelled many East-German researchers from their jobs, resulting in a sizeable wave of job mobility, from the West to the East, and (less so) from the East to the West (Meske 1993; Günther et al. 2010). I sketch these years here, as the writing of the programmatic text "Measuring the knowledge base" cannot be understood without being aware of these circumstances, and also because Loet's writing was always rooted in the analysis of world as it presents itself. Next to the disappearance of the socialism in Europe, both globalization and the emergence of the internet characterize this era.

---

[4] The term *Wende* is German and its literal translation is *turn*. It is usually applied to describe the period between November 1989 (opening of the state border of the GDR in Berlin) to October 1990 (dissolution of the GDR and integration into the Federal Republic of Germany).

'The Measuring the knowledge base' consequently starts from the observation of the importance of science (in the broadest sense) for knowledge-based economies. Knowledge-based economies and the innovation taking place within them were the guarantee for social market economies and a just, inclusive welfare state – of this, Loet was deeply convinced. So, to contribute to a better world, one should aim to study this world in a way that can identify the conditions which foster innovation. Coming from a Marxist-Leninist philosophical background myself, where I learned that society ought to be analyzed scientifically, the holistic, all-encompassing view Loet aspired to resonated very well with my thinking. I was rather baffled sometimes by the diversity and scattering of various discourses in the West. My old institute, the previously mentioned ITW, housed departments on Innovation Studies, History of Science, Sociology of Sciences, Science and Technology Studies, and Philosophy of Science, <u>all under one roof</u>. In addition to being brought up in an intellectual environment which fostered an inclusive view on science studies (*Wissenschaftsforschung*[5]), my first academic background was in theoretical physics, more specifically thermodynamics of irreversible processes and physics of self-organisation and evolution (Haken et al. 2016). The former physicist (Andrea) and the former chemist (Loet) also met epistemologically.

The text "Measuring the knowledge base" moves from the problem analysis to the dimensions of future studies. The emphasis was on the role of networks for innovation, the emergence of innovation at interfaces between various structures and the role of niches.

> "Under the condition of globalization, local niches can gradually be dissolved because new horizons offer other options. As the relative weights of relations in a network change by ongoing processes of collaboration, appropriation, and competition, the new balances and inbalances can be expected to generate a feedback in the knowledge infrastructure at other ends. The (sub)systems can then be expected to recombine into new solutions with degrees of success." (Leydesdorff, Scharnhorst 2003, p.4)

As this paper will exemplify, newly emerging structures can take the form of new institutions and but also that of new practices. Both come with new forms of communication, sometimes by recombining existing ones. In any case, these communications lead to new form of traces and can open up new ways of measurement. At the same time, as expressed in the 2002 text, one has to be cautious both conceptually and methodologically when dealing with possible new forms of measurement.

---

[5] As an example of topics done under this term, see the *Gesellschaft für Wissenschaftsforschung* and its yearbooks (http://www.wissenschaftsforschung.de/Index.html, Permalink: https://web.archive.org/web/20240909224243/http://www.wissenschaftsforschung.de/Index.html).

> "The windows for studying subjects as intangible as knowledge production and communication, have to be carefully reflected as the order of communications is not "naturally" given. We are constructing second-order constructs about knowledge-based constructs." (Leydesdorff, Scharnhorst 2003, p.12)

This is why including philosophy in quantitative studies of science is not a *nice-to-have*, but imperative.

I have to admit, that while Loet and I discussed the ideas behind a programme as proposed in "Measuring the knowledge base" a lot, the actual writing of the final version has to be credited almost entirely to Loet. In 2002, when the deadline of submission was near, my just-found new academic home at the Royal Netherlands Academy of Arts and Sciences (KNAW), in the form of the inspiring NERDI group (Networked Research and Digital Information) and located at the NIWI (Netherlands Institute for Scientific Information Services)[6] (Blauw 2005), dissolved – or better yet, was *abgewickelt*. This was nothing short of a personal crisis for myself, as I had just believed to have found stability after the traumatic years of *Abwicklung* in the years of breaking up the GDR society (economy and academia included). In those years, the former employees of the GDR Academy experienced annual evaluations – and even if positive to no prevail, and eventually short-term, often half-time project jobs funded by the so-called Researcher-Integration-Programme[7]. Reliving another *Abwicklung* inhibited my work abilities in this period, and Loet became the first and leading author on the final text.

But, writing this paper now, I realize once more, how much the text 'Measuring the knowledge base' represents the foundation of the discussions Loet and I had which continued over decades. Even if this text did not produce any new project money, many other follow-up projects did. We kept regular meetings which often started with a dinner and ended at Loet's house, and doing so we returned in essence to the research lines laid out in the original text. This is why I'm convinced that the three cases I will lay out, covering a period of 10 years of research, are very much informed by the programme we set out at the beginning of my Amsterdam life. However, before coming to the cases, I will first consider some wider aspects of quantitative studies which also have been addressed in the 2002 text.

## Why do we need to measure?

The idea behind the text "Measuring the knowledge base" was to bridge the conceptual views on innovation and concrete empirical studies. The image on the title page (Figure 1) shows the Triple-Helix dimensions and layers of knowledge production in society together with data and indicators, as imagined 20 years ago. Looking at this illustration today, one can say one thing for sure: We have much more data. In this paper I will then ask: But do we also have more knowledge? This will be the red thread weaving through this text.

---

[6] https://nl.wikipedia.org/wiki/Nederlands_Instituut_voor_Wetenschappelijke_Informatiediensten
[7] And at least we had this: https://de.wikipedia.org/wiki/Wissenschaftler-Integrations-Programm#:~:text=Das%20Wissenschaftler%2DIntegrations%2DProgramm%20(,%2C%20Sachsen%2DAnhalt%20und%20Thüringen .

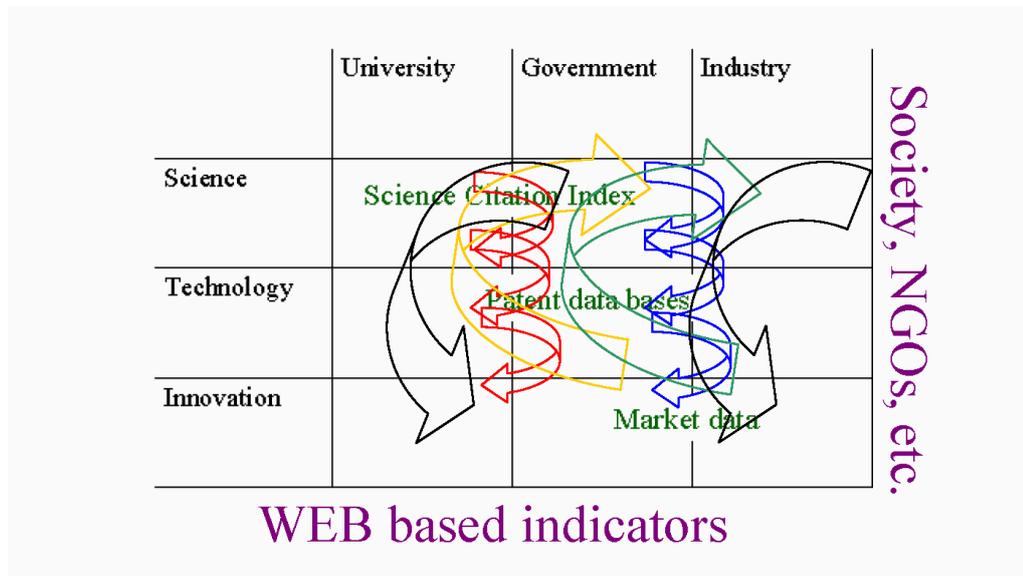

Figure 1: Illustration at the title page of the text 'Measuring the knowledge base'

As mentioned in the introduction, this paper is based on a presentation that was given at the workshop, "Philosophy of Science Meets Quantitative Studies of Science"[8]. This then explains the philosophical (rather than scientometric) lens that I will take. Indeed, nothing could be more suitable to honour Loet Leydesdorff, who always – even in writing a funding programme text – excelled in *being philosophical* (meant here thoroughly reflexive). This holds for all his studies, even for the most meticulous and methodological heavy ones.

With the success of the field *Quantitative Studies of Science,* an observer could say that the study of the sciences has moved from the reflective philosophical corner to the rather (natural) science corner; one could even say, it has moved from being a gentleman's amusement within science history and philosophy to become an instrument for managerial need. I often start lectures on aspects of scientometrics and indicator research with the following narrative: The science system grows, and after the Second World War it almost exploded. As there are public funds going into it, monitoring is needed.

Interestingly, I as others see the same trends in the West as well as in the East (Wouters 1999). If one reads – as an example – the Essays of an Information Scientists of Eugene Garfield (Garfield 1977-1993), you can feel the excitement about all those new possibilities to 'measure'. By the way, and this is something I feel increasingly touching, this innovative push came with a very modest, carefully testing-the-grounds attitude about what machines or what automation might be able to do for us. The main focus was primarily, at this time, on how to best find information to foster science, but also how to understand academia. The latter is visible in the insights produced by a network of researchers such as Merton, De Solla Price, Lederberg, Garfield and many others (Elkana et al. 1978). I think the makers of these new information systems were themselves taken a bit by surprise later by the adamant use of those new information services for evaluation. The early versions of the National Science Foundation Science&Innovation indicators (NSB 1993) breathe the same spirit: numbers one needs to have in order to steer the sciences to ever growing progress. In the East, we find a similar development. The famous book of Nalimov and Mulchenko (1969) was also born out of the need to manage the science system which increasingly became complex.

---

[8] Information about the workshop can be found here: https://www.llc.unito.it/eventi/workshop-philosophy-science-meets-quantitative-studies-science (Permalink created July 2, 2024 https://web.archive.org/web/20240702141951/https://www.llc.unito.it/eventi/workshop-philosophy-science-meets-quantitative-studies-science ). The presentations have been recorded on YouTube https://www.youtube.com/playlist?list=PLPlZ4NrhUoUQYD74Qz2bnmK18YqqaKaPB, and slides are available here: https://drive.google.com/drive/folders/1oSZlHU8KaRfa3MCS0VkdU25T4pzDGKXc

Automatization created both the means and the need to further develop quantitative studies; the emergence of the internet as the last big wave of automatization in the 1990's has only amplified these trends. It remains to be seen how the 'new generation of machines' (generative Artificial Intelligence) will again shake up the situation within science studies. So, in short, one could summarize: In the digital age we are all ruled by numbers. The text 'Measuring the knowledge base' acknowledges this fact but also asks to look beyond numbers.

Scholars may be seduced by the fact that we <u>can</u> count, and academia in general operates also under the imparative that we <u>should</u> count, for all kind of reasons, mostly in the accountability of spending tax money. In this regard, we often seem to have forgotten the limitations of counting, for which we actually need to consult philosophy. Seeking truth in numbers represents a positivistic approach and we, sometimes I think, have forgotten about the meaning behind numbers, something that Loet deeply cared about.

> "The observations and indicators are also knowledge-intensive, as one can no longer assume that the data is readily at hand. The overwhelming availability of information in a knowledge-based society makes it necessary to reflect on the selection of data from a theoretical perspective." (Leydesdorff, Scharnhorst 2003, p. 10)

## How to measure emergent structures?

In the next sections, I will share some lessons I learned during my career and insights which I gained, all of them centred around one of the problems identified in the "Measuring the knowledge base" text. With the choice of an autobiographical compass, inevitably there comes an element of arbitrariness, I must admit. Therefore, I will briefly discuss three examples, all of which are situated in structures which act as intermediaries. They are also a bit different in nature, and for all of them I have chosen to discuss some measurement challenges.

In the text 'Measuring the knowledge base' the aspiration was to come up with 'new innovation studies' and relate to them 'new indicators' which were able to shed light on newly emerging structures in the organization of the sciences. By emergent I mean structures which have not existed before and which appear due to self-organising processes in academia (Scharnhorst et al. 2012). Loet and I were deeply convinced that only an evolutionary approach would do justice to identifying the knowledge base and how it changes. Some parts of Loet's research in the last two decades was devoted to the exploration of new scientific fields that were born out of interdisciplinary work and in turn become manifested in new journals. These new fields can be traced by following topological changes in the citation networks among journals (Leydesdorff, Zhou 2007). However, formal scholarly communication is only one of the many possible traces for innovation. Loet was hunting for the 'right unit,' where innovation manifests itself – be it geographically in countries or regions (Leydesdorff, Cucco 2019) or communication-wise in patents or articles (Leydesdorff et al 2015). In "Measuring the knowledge base" we argue:

> "The uncertain definition of the unit of analysis for studying a knowledge-based system of innovation in terms of nations, sectors, technologies, regions, etc., brings in new players as potentially important contributors. …. " (Leydesdorff, Scharnhorst 2003, p.6) Only to later state that "The network overlay emerges as a new unit of evolution. When this structural innovation can be temporarily stabilized, it may begin to coevolve with the subdynamics upon which it builds." (Leydesdorff, Scharnhorst 2003, p.10)

In the text of 2002, we point to networks as a candidate for such a (new) unit of analysis. The examples I have chosen for this section are located at the interface between research and research infrastructures. My own career path led me to work in and for research infrastructures – from NIWI (Blauw 2005) (an institute build around the former library of the KNAW), to DANS (Data Archiving and Networked Services, a data service provider) and DARIAH (a European Research Infrastructure Consortium) (Morselli 2023). My examples are rooted in my experiences working in those institutions. They concern: the DANS archival service; the Key Performance Indicator discussion in DARIAH; and, most recent, a new research management approach developed in an EC funded research project – the Polifonia Research Ecosystem. In each of these cases, network aspects play a role. DANS as an institute is part of a wider landscape of Research Data service providers which operate in a networked way. DARIAH is essentially is a network structure bringing together national research infrastructural nodes around a certain methodological innovation. The Research Ecosystem approach of Polifonia is based on the networked character of research management in interdisciplinary projects.

### How to measure a Data Archive?

I entered DANS in 2012. At this time, DANS was a relatively new KNAW institute, founded in 2005 around the first web-based service for researchers to self-archive their research data (prior to the Dataverse Network which started 2006, FigShare in 2011 or Zenodo in 2013). One could say, DANS represents itself an institutional innovation albeit being situated inside an established institution such as the Royal Academy.

Within DANS, we set up a research group. This was somewhat experimental as DANS is clearly an infrastructural, research-supporting institution, which I still tend to call a data archive. However, the portfolio of DANS is wider than 'just' being a data archive. Usually, research institutions and research support institutions don't mix, although the boundary between practices in either of them is always blurry (refer here to the discussion on so-called Knowledge Infrastructures (Edwards et al. 2013)).

In one of the projects at DANS (Impact-EV[9]) we had resources and in-house capacities to look into information registered around our digital data archiving services. This coincided with the visits of Christine Borgman[10] at DANS in the period of 2013-2015, funded via a KNAW Visiting Professors Programme. Christine arrived with all her background on ethnographically based information studies on data uses, data practices embodied in the "Center for Knowledge Infrastructures (CKI)" which she founded in the UCLA Department of Information Studies, and she now wanted to do a project with us at DANS. Within the same period, Andrew Treloar and Herbert van de Sompel[11] also visited DANS. They were both very intrigued by the possibilities of web-based machine-readable scholarly information as discussed in the community around the *Article of the Future* (Bourne at al. 2011). The possibility to make scientific knowledge production traceable, and so better reproducible, intrigued and intrigues many. So, for DANS services itself the question emerged: Can we actually trace what the people who interact with our data services do?

This is not an absurd question. Their interaction with our data services is web-based, of course, and so we have log files. I thought, being a statistical physicist by first training, 'oh that's actually great, we have for sure lots of data' – a bit naively as it turned out. But, we could do and did a lot, as is visible from Figure 2 (Borgman et al. 2015). For instance, we looked into the geographic distribution of user visits, and found that we also have (non-robots) visits from outside The Netherlands and Europe, despite being a national data archive. We mapped the collection items from their catalogue entries and visualised the disciplinary distribution of datasets. We looked into distributions of file sizes, files per dataset, downloads per datasets, etc and found the skew distributions characteristic for any complex system. We analysed the technical metadata of the datafiles, identifying main used format types, which has implications for the long-term curation of data. We were even able to see in the logfiles, which keywords where typically used together.

---

[9] Impact-EV "Evaluating the impact and outcomes of European SSH research" funded by the FP7-SSH - Specific Programme "Cooperation": Socio-economic Sciences and Humanities. Running 2014-2017. Grant PID 613202
[10] https://christineborgman.info
[11] https://hvdsomp.info and https://andrew.treloar.net

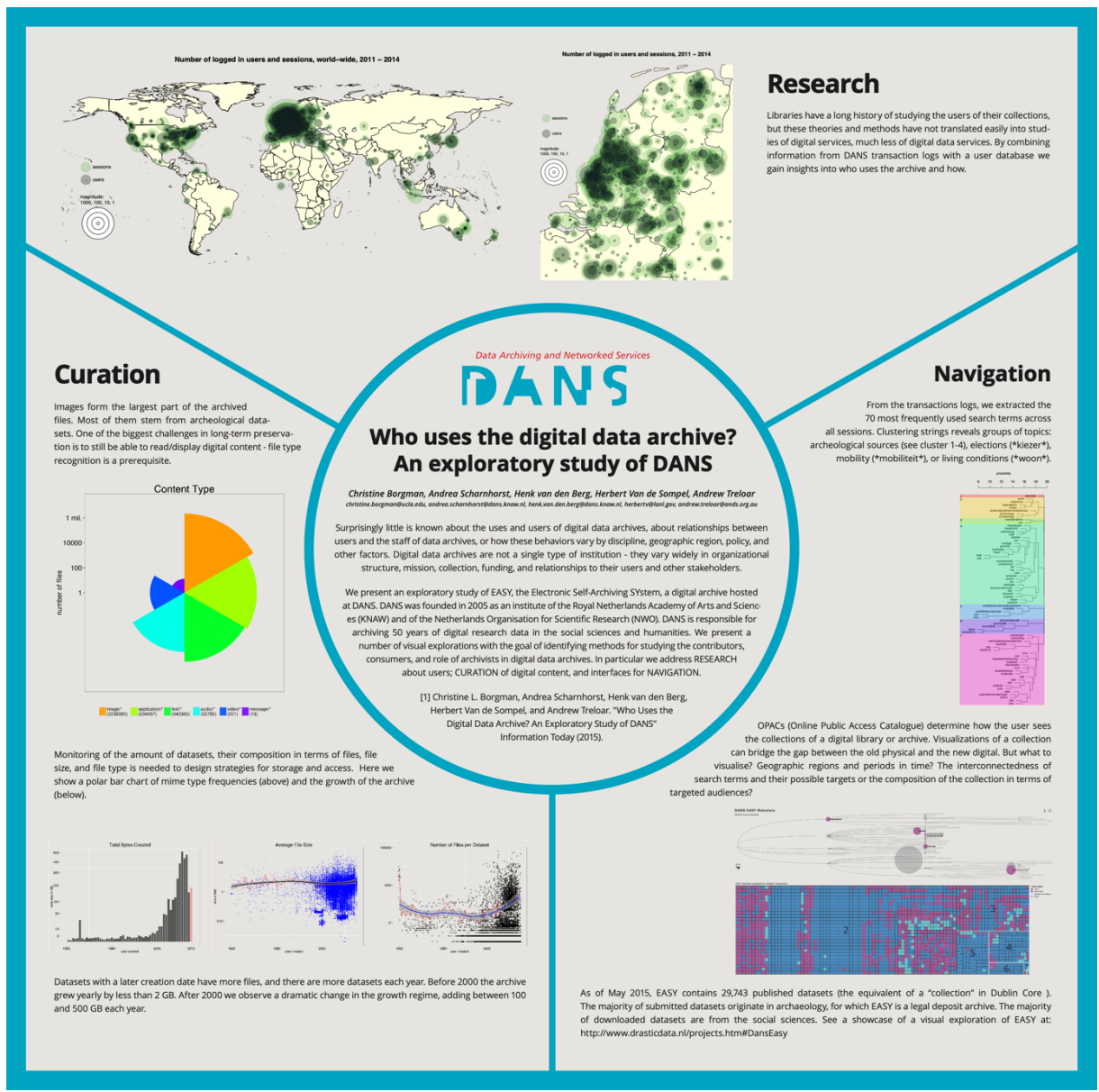

Figure 2: Poster for ASIS&T 2016 – larger image can be found here (Borgman et al. 2024)

While proud of our quantitative analysis, I want to emphasise that the main insights did not come from the quantitative explorations but rather from the qualitative study conducted by Christine Borgman, her students, and me (Borgman et al., 2019). To this qualitative study, it was the quantitative studies that provided necessary information, enabled us to find the interviewees, and enabled Christine (and the wider research team) to get a grip on size, structure and dynamics around the archive. What we learned from the qualitative studies were things like motives and ways of people to engage with data archives – nowadays, this is called *data practices* or *data culture* (Borgman 2015, Borgman et al. 2019). These were insights which could not be obtained by the quantitative approach.

When it comes to measurement, the lesson that I learned was the following: Yes, in (automated) information services, a lot of data are recorded for specific reasons. We could see traces of users but we could never properly follow one of their journeys, let alone the journeys of a whole group of users.  So, to summarise, we performed some IP analysis (see the geo map in Figure 2), we did some query analysis and found clusters of search terms, and we did the bread-and-butter statistical analysis delivering the usual skewed distributions in file sizes and files per datasets. However, we could not do what I had hoped for – a tracing of actions which leads to the various types of interaction with the archive.

The reasons for this are simple: The log-filing was tailored in a way to let the machinery run smoothly, with the mindset of how to run the archival service in order to manage visits and uploads and to keep the provenance around the technical metadata. In other words, it was optimized in order to monitor the performance of the IT system; not to monitor its actual use. One could compare this to tracing a mundane infrastructure. Think here about a railroad system. There is information about the positions of trains (and this was the type of information we found) but not of the passenger streams around them.

As an almost ironic footnote, that we were able to trace some actions of individual users had to do with a specific feature of the data archival service at this time. To download files from the data archive, an account was required, enabling the tracking of individual accounts. With the more recent turn to Open Science, accounts are no longer required to have access to datasets. The metadata of our datasets were always open, but now the majority of data sets themselves also are open. So, the (limited) tracing of users that was possible in the recent past is no longer a given and openness here also leads to less information.

The take-away message from this first case is thus: do not place too high hopes on numbers and certainly do not believe too much in them. ***Often, we have a lot of numbers – but most often we don't have the numbers we need.***

We really need to remind ourselves now and then on these limitations of quantification. I'm still somewhat baffled when I see a new information system popping up or being set up, and there is endless optimism regarding the accuracy and utility of possible metrics that can be collected. This specific row of measuring experiments on the usage of data held by a digital research data archive archive also added a layer to what I now personally think about how much data in principle can become re-purposed.

Regarding the practical aspect of this case at hand and the research questions we wanted to ask back in the midst of the 2010s, I still defend the need to look into measurements around research infrastructures such as libraries and archives. Yes, we need a kind of *archivemetrics* (Dorsch 2023). As in the history of quantitative studies of science, measurement around collections has been very informative – think here Bradford's law (Mayr 2009) – and so is any kind of metrics around information systems (Priem et al. 2010). Therefore, I think we can and should do quantitative studies also around data practices and the research infrastructures, but we should primarily clarify the information recording structure before we rely on quantitative studies executed in it. Quantitative studies as an exploratory method can help to identify limitations of those new information structures.

A good example here are recent data citation studies. These studies are more readily available because data citation indices have been set up in a fashion similar to the article citation indices. But the related standardized data citation practices are very 'young' (Gregory et al. 2023). If we recall that the referencing behaviour in formal scholarly communication goes back centuries to references in letters, even before journals appeared; formal data citation is in its infancy. In other words: the science citation index started on the basis of an already established broadly standardized reference culture to journal articles. Often, data citation studies deal with very messy collections where the interpretation of the results delivered requires extra care, as standardization is just about to be implemented and not yet fully sorted out. In extending the lesson formulated above about missing the right numbers, one could also say: we are recording a lot in the digital age, but we often don't record what we want to have in the future.

When writing the text 'Measuring the knowledge base' we were very much aware of the need for a good data collection from the start.

> "Systematic data collection, however, requires standards. The matching between the analytically relevant questions and the institutionalized routines asks for an informed trade-off between considerations of a potentially very different nature. How does one define a baseline? How does one normalize? What is/are the relevant system(s) of reference? Scientometric indicators cannot simply be "applied" in another context without generating terrible confusion. Scientometrics is a research effort in its own right, since the indicators have to be reflected." (Leydesdorff, Scharnhorst 2003, p.14)

### How to measure an ERIC?

The 'Measuring the knowledge base' text looked for new indicators, new measurement approaches related to newly emerging structures in academia. My second case about such new structures and their possible measurements again comes from the autobiographical compass I chose to unfold cases. It concerns a research infrastructure network with which I'm now working for eight years, the DARIAH European Research Infrastructure Consortium (DARIAH ERIC). DARIAH stands for Digital Research Infrastructure for the Arts and Humanities. DARIAH has existed for 10 years now (since 2014), and is part of the European Strategy Forum on Research Infrastructures (ESFRI)[12] roadmap, listed there as a landmark research infrastructure.

Some readers may be unaware that an ERIC is actually a legal construct, an organization registered in one of European Union member states and running on membership fees of committed EU countries. This form of coordination was introduced by the European Commission to align infrastructural efforts in the various European member states. The EU knows various of such cross-national initiatives – some of them set up as projects, some as funding streams, and some as new institutions. The European Open Science Cloud[13] is another such example. Scientists, and even science policy experts too, often have a hard time to understand all those emerging intermediary layers.

Coming back to DARIAH, it is important to reiterate that an ERIC is a legal organization, not a project. I was initially brought into the administration of DARIAH during a phase when DARIAH was transforming from a funded project into an organisation. All parties involved worked together to set up a governance structure with new administrative processes, sorting out the *modi operandi*.

---

[12] According to its own website, ESFRI is a strategic instrument to develop the scientific integration of Europe and to strengthen its international outreach. https://www.esfri.eu/about
[13] EOSC Association https://eosc.eu

DARIAH and other European Research Infrastructures come together in a network called ERIC FORUM[14], and in this network a debate started several years ago about so-called Key Performance Indicators (KPIs). I recall how the DARIAH president at this time, Jennifer Edmond, came back from one of these meetings, and shared her experience. Essentially, representatives of the various ERICs (representing various scientific domains) were discussing a variety of issues, and when it comes to KPIs, they only talk about publications and impact factor, Jennifer reported. We (at DARIAH) were all aware that such a measurement approach would not be adequate for the humanities. A large part of the humanities doesn't rely on the journal model for scholarly communication. So, any statistics based on such data collection would represent the humanities field very poorly. But, we also agreed that we cannot say: 'no KPI's for us please, we are too special'.

Jennifer is an innovative humanities scholar; during her term in DARIAH as president, she was a dedicated and vigorous research manager. At this time, we had recently run brainstorms on missions, written a strategy document (DARIAH 2019a) and developed a strategy-implementation plan (DARIAH 2019b). We had also discussed how we could find evidence for the benefits we were convinced DARIAH was bringing to the Digital Humanities communities in different countries. In this case, we started from the question what could be counted meaningfully, not – as in the previous DANS case – what has already been counted.

Of course, we also looked into what information we already had. We knew what we wanted to measure, but those numbers were not immediately at hand. We found a solution to this problem by combining existing sources and establishing new data collections. We relied on two sources of information that DARIAH collects: first, the services and activities the national nodes had to report (in-kind contributions), and second, the output, activities and service use we could measure on centrally shared platforms. Eventually, a reporting system for every country node in the ERIC was created which is fed from various sources. While not necessarily an elegant solution, this system combines publications (the traditional sources) with information on services and network activities. Technically, we partly rely on Google Scholar, Zenodo communities, Zotero libraries, the SSH Marketplace catalogue, and also Google Sheets, etc..

The key goal of the reporting system is to deliver ways to express the additional benefit which comes by operating as part of a network. In the DARIAH Strategy, we wrote "what we really want our KPI's to provide evidence for and focus our efforts upon is the depth and richness of our impact into research communities, into national consortia, into practices and knowledge base of individual researchers." (DARIAH 2019a, p22). For example, events (participation and organization) are part of the indicator set. But, there is no ready-made database in which we can just query. Therefore, for the event-related indicators in this case, we came up with our own documentation system. We classify events according to audience, size, national/international importance, and specify our role in them. From this, we can produce infographics of the kind shown in Figure 3, which in turn we monitor annually and publish in the DARIAH Annual Report (DARIAH 2022).

---

[14] https://www.eric-forum.eu

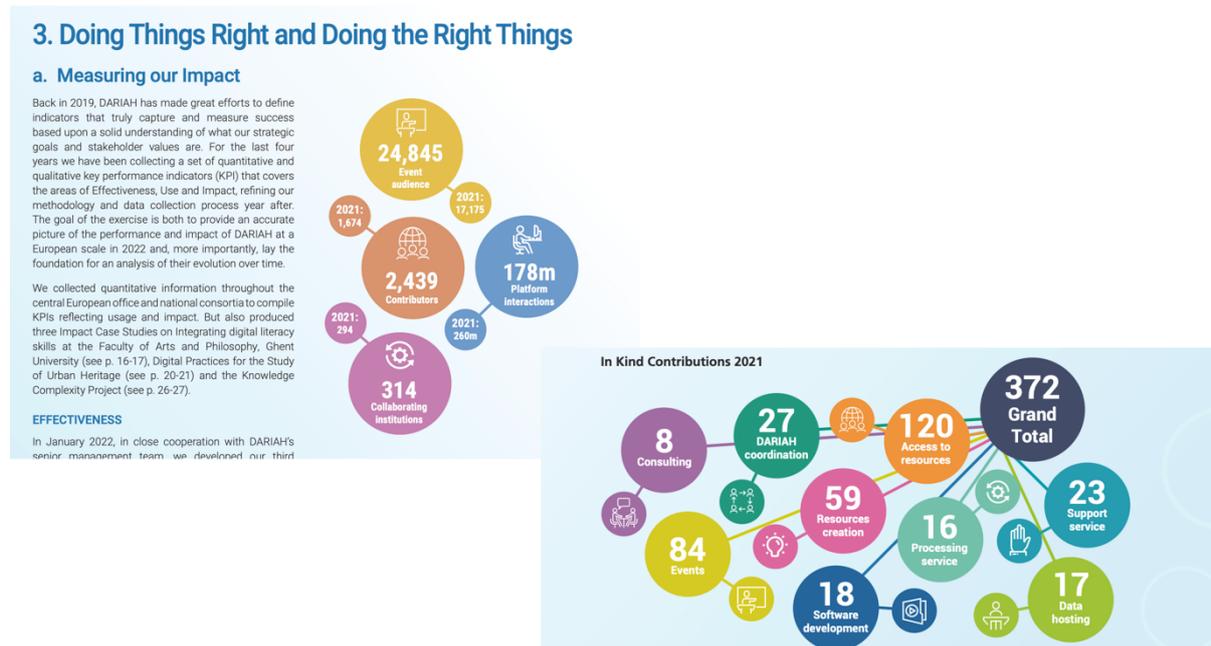

Figure 3: Visuals from the DARIAH Annual report 2022 (DARIAH 2022).

Reading back the text 'Measuring the knowledge base', I was surprised to see that Loet and I had intellectually anticipated some of the dilemma's which come with newly emerging network functions.

> "The network overlay emerges as a new unit of evolution. … Given the selection pressure of the new dimension, old institutional arrangements may survive, but will probably have to adapt their function, as well as their form, to the new environment. The hypothesis of a "global agent" can be formulated as the expectation of change because of the selection pressure on institutional arrangements. The global agent, however, remains a network function and consequently operates as a regime of uncertain expectations." (Leydesdorff, Scharnhorst 2003, p.10)

Perhaps I have worked too long in the area of quantitative studies and have to confess, I no longer believe numbers, or at least, not too readily. I'm also intimate enough with the process of information gathering, being a content-generating user myself. So, looking at the DARIAH KPIs, I know that we do our best and produce decent infographics out of the practice. I also think we make a good argument in favour of counting more than just publications. However, if you really want to talk about the impact of infrastructures, I'm not convinced these numbers are the best evidence that you can find.

Impact is something everybody is after but is also notoriously hard to measure (for a review about impact measurement in the social sciences and humanities see (Reale et al. 2017)). For sure, we forget half of the things we actually have an impact on. This is partly because we as individuals operate in various contexts, within various networks and we have different hats that we wear, so to speak. These networks often influence each other but they all also require some kind of distinct accountabilities; we sometimes forget that when we go somewhere in one role, we are also contributing to other networks or fulfilling other roles.

Surely, if this is the case, we should get this documented at various places. But then as researchers, we are also aware that we create double counts, which does not contribute to making a metrics more reliable. So, even the numbers collected in the first place come with some error margin. I dare to say then that the best of these measurements or indicators we came up with are not the actual numbers (the values of the variables) but the indicators (variables or categories) themselves – the important dimensions and, at best, the order of magnitude these measurements take.

So, the lesson learned here is: **the really important things are difficult to count and be counted**. It is not impossible, but each metrics develops its own dynamics, a life of its own and on-going **reflexivity in indicators creation and implementation is needed** (Biagioli, Lippman 2020, Goodhart 1975).

> In the text 'Measuring the knowledge base' we formulated "However, the focus on knowledge-intensive developments requires us to take a reflexive turn towards the data gathering process, both in the quantitative and in the qualitative domain. The program of innovation studies is anti-positivistic, as one begins with expectations instead of the observable "facts."" (Leydesdorff, Scharnhorst 2003, p. 12)

Alongside expectations and reflexivity, indicators rely on documentation. However, and this brings me to the last case, creating documentation is a skill in and of itself. This is a fact which is almost forgotten. It may be that the seamlessness of current IT systems, the apparent ease with which anybody can fill a research information system with content, leads us to think that the right information ends up at the right place almost magically. Now, whoever has documented her own output and activities in a database should be aware that there are always many options in bibliographic documentation. Even with standards, there is always a possibility of erroneously providing content under the wrong category. To think that researchers can be their best own documentalists is an illusion. Still, they are obliged to do so – even more so under the current manifestation of Open Science.

### How to measure a project?

The last case in this section about the measurement of emergent structures concerns a practice rather than an institution. The project-driven nature of current research management comes with requests to engage in various forms of documentation. Among them are financial and technical reports, deliverables and, as a relative new element, Research Data Management Plans. The consortia formed around funded projects are networks that are much more ephemeral than the data service provider or the ERIC I discussed in the previous sections. The projects themselves are not institutionalised, but a lot of their execution follows nowadays standard routines. This holds in particular for their management and related well defined documentation forms, think here reports as deliverables. Documentation is needed both for the sake of the accountability of the grants received, but also to provide evidence to reviewing experts about the scientific goals achieved with this funding.

When I talk about 'how to measure a project,' I want to present a specific case and a specific type of documentation, namely the Research Data Management. The project in question,

Polifonia[15], took a slightly new approach on the request to monitor the use of data in the research process, or in other words, the Research Data Management (RDM) and its documentation in the form of a plan.

Polifonia (2020-2024) was coordinated by Valentina Presutti of the University of Bologna. It brought together semantic web technologies and cultural heritage. In its own words on the website, the project aspires to play the soundtrack of our history. It brings together various types of information such as about the soundscape of Italian historical bells or the influence of French operas on traditional Dutch music. The project departs from the belief that European cultural heritage hides a goldmine of unknown encounters, influences and practices that can transport us to experience the past, understand the music we love, and imagine the soundtrack of our future.

One requirement of the funding stream in which Polifonia was granted was to use a specific template, the so-called Open Research Data Pilot (ORDP), to document its Research Data Management. This template guides the provision of information around data by detailed questions. Research Data Management Planning is increasingly standardised, including platforms and tools which help researchers to fill in questionnaires and other templates around data.[16]

But, believe me, researchers do not know their data. No this did not come out right – of course they *know* their data, but they often don't know how to *document* their data. While some scientific communities have standards and long-established practices on how to make their data FAIR (Findable, Accessible, Interoperable and Reusable) (Wilkinson et al. 2016), some do not. Where to enter information about metadata schemes, metadata content itself (such as file formats) can be very confusing. It becomes even more complex in interdisciplinary projects where different data documentation practices collide. This is the reason why good Research Data Management (RDM) is best done in a collaboration between data stewards and researchers, both able to contribute their slightly different skill sets.

The attention for RDM represents a certain revival of documentation. It is also important to notice that the usual documentation during research processes and a more formal documentation of certain aspects of research (think here referencing publications or data) are often executed based on different mindsets. Research emphasises the fluidity of thought, as this is at the heart of any investigation and exploration. It is an iterative operation which often starts with uncertainty and in interdisciplinary projects necessarily also with ambiguity (Galison 1997). Documentation can likewise be iterative, but it should be meticulous and firm. In the case of RDM, in each RDM plan there is most likely a table with data sources described as precise as possible. The documentation needs to be consistent to meet Open Science requirements.

---

[15] Polifonia: a digital harmoniser for musical heritage knowledge funded by the H2020-EU.3.6. - SOCIETAL CHALLENGES - Europe In A Changing World - Inclusive, Innovative And Reflective Societies programme. Running 2021-2024. Grant PID 101004746. See: https://polifonia-project.eu for the project website (Permalink https://web.archive.org/web/20240704124553/https://polifonia-project.eu/)

[16] See for instance the tools listed at the GO-FAIR website, subsite RDM starter kit https://www.go-fair.org/resources/rdm-starter-kit/ Permalink: https://web.archive.org/web/20240711165033/https://www.go-fair.org/resources/rdm-starter-kit/

Returning to the knowledge production process in a project, in the beginning, researchers often don't know which of the data they have in mind they will actually work with. Initially planned data sources may be discarded as new data needs emerge or new data sources become available. Research is an intrinsically messy process (Wouters et al. 2008), and data handling in research is equally so (Gregory 2021). This makes data re-use a challenging norm (Borgman, Groth 2024). This might also be the reason why RDM is considered such a cumbersome exercise – often more seen as an externally defined 'must' than something which really supports research.

The more functional approach Polifonia took for RDM (and which was labeled as *Ecosystem*) actually emerged from a research need. The Polifonia project was driven by semantic web specialists, and so Linked Data solutions formed the methodological ground when it came to the technological backbone of the project. These new Linked Data solutions (ontologies and knowledge graphs) concern new data sources around musical objects such as sound, scores, as well as literature about musical events or detailed documentation of musical instruments such as church bells or organs. Collective software development was essential for the project. The main motivation to set up an *Ecosystem* came from the insight that 'it would have been impossible to develop a single framework incorporating all the technical outputs in a single end-user tool. Music is such a heterogeneus domain that the project consortium knew they could not promise that. The Ecosystem idea was developed because of the need to give a coherent presentation of a collection of diverse outputs. The connection with Data Management was natural but arrived afterwards'[17]. As we will see later, in a kind of second order of operation, this group of researchers applied their computer science and engineering skills not only in the research itself but also for its documentation. (Daga et al. 2023)

By providing resources for a data steward among other means (such as a Technical Board) to guide the process of data management, Polifonia's aspiration was able to combine the management of data and the management of the project in a way which supported both: the actual research and its documentation. Before I describe the Polifonia *solution*, I first want to sketch the project more clearly, as this context also explains the efforts taken for documentation.

In the proposal phase of the project, the proposal writers were already aware that this project tackled a lot of problems at the same time. Concrete musicological questions such as 'what is a melody?' and 'how to compare melodies?', met questions such as 'which interfaces would what kind of archetypical user need to have in order to become enticed to interact with music?', or 'how can a user be supported in serendipitous search?'. On the overall informational side of it, a single question stands out: How can musical objects recorded in forms as different as sound, images, and texts be documented in a way, which respects the various views or perspectives of scientific specialties on them and still bring them all together in one documentation system? Specialty views are usually represented by classifications tailored towards specific collections. How could an ontology for such a web of musical objects be designed which is aligned with existing standards, but also gives room for new ways of representing objects?

---

[17] Personal communication Enrico Daga

In the project layout, the consortium chose to complement the usual work package structure with 10 pilots. Those pilots formed the 'use cases': they defined the requirements for computer engineering solutions, they interconnected the work packages, and they also were the practice test for new solutions. As can be seen in Figure 4, pilots are grouped around different types of activities: from more infrastructural ones such as *preserving* and *managing* over research specific ones such as *studying* to wider outreach such as *interacting*.

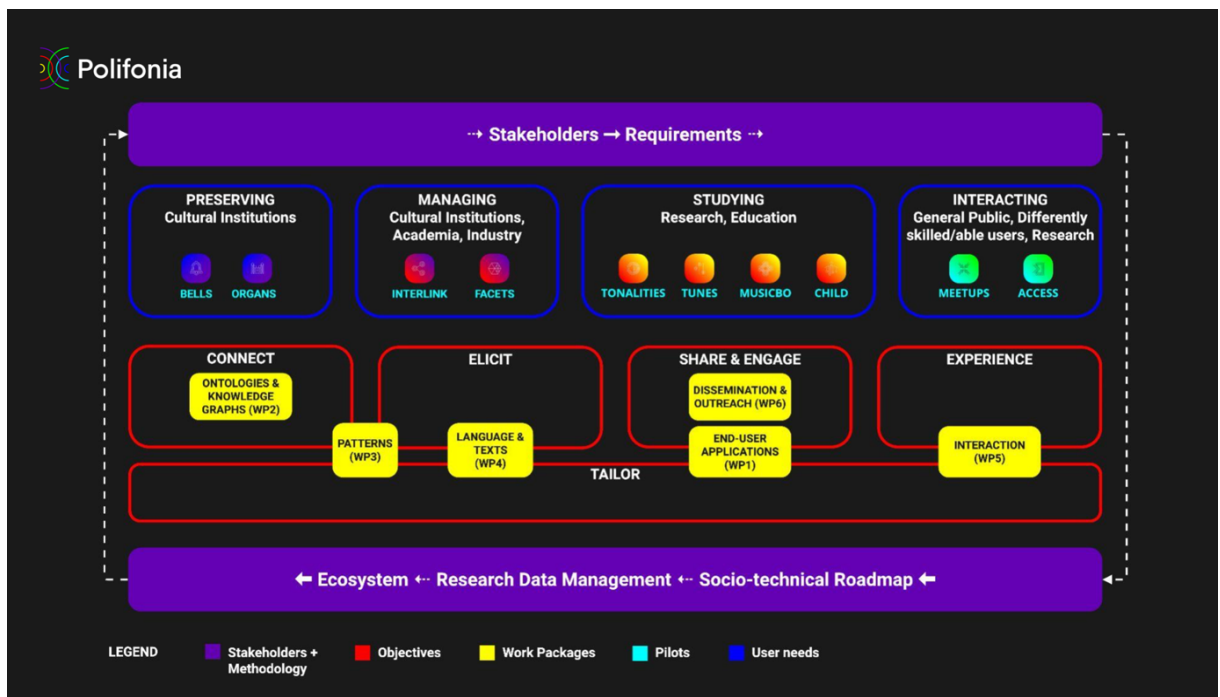

Figure 4: Overview about the Polifonia project – slide courtesy of Valentina Presutti and the Polifonia Consortium (produced on occasion of the final project reporting May 2024)

This 'pilot approach' served various functions. With each pilot, a niche was created to explore specific problems in an independent way. At the same time, the project provided means to bundle pilots together to harvest synergies. As shown in Figure 4, this was achieved by commonly working on requirements (also with stakeholders) and a firm planning according to a socio-technical roadmap. The construction of the socio-technical roadmap was based on surveys which brought together epistemic questions (what to research and why) and methodological choices (how) with information about the objects of study. Hereby, the main components of what was called the 'Polifonia Ecosystem' were defined. The 'Polifonia Ecosystem' information system formed the basis for the Research Data Management of the project.

As stated above, the project relied on collaborative software development. More specifically, the production of machine-readable data hubs (knowledge graphs) using semantic web methods was a requirement shared across all pilots. The decision was made to use GitHub as shared platform to document all software development. ' The methodology allowed to gather metadata directly from the developers in-situ – meaning in the same place where engineers work.'[18] So, with hindsight it seems only natural to also have used GitHub for the monitoring of the main assets or components of the Polifonia research process. But, as always, the actual process of setting up the documentation was much more complex. One should be aware that managing software production (using the GitHub features) and managing the main components in a research process come with different requirements.

The Polifonia Ecosystem required a design process which began with blueprints already contained within the proposal. The implementation and iterations to adapt its structure was supervised in monthly meetings of the Technical Board. Eventually, workshops were organised to curate content for the Ecosystem. This information in turn was finally harvested for the Research Data Management Plan (Scharnhorst et al. 2023) and for a reusability study (Carvalho et al. 2024).

From an information science perspective, the Polifonia Ecosystem (Daga et al. 2023) occupies a middle ground between the bread-and-butter usual project timelines and work packages interdependencies and the fine-grain tracing of single research objects or workflows around them as discussed in the context of Open Science. At the core of defining such a Research Ecosystem, one defines a selection process from which all intermediary products in the research process, those which are constitutional for the project, are determined. They are then called *components* and are classified in categories such as *Data, Tools,* and *Reports*. Components are those research elements or assets which needed the closest monitoring. Obviously, this selection is specific for each project, and might also change during a project depending of the course of the knowledge production process itself.

In the case of the Polifonia Ecosystem, for each component a *champion* (responsible programmer) was defined. Each component has a representation in GitHub in the form of a GitHub repository. This repository can be part of the Polifonia GitHub community, but it can also exist outside of it, therefore enabling all contributors to build on former work. As long as the component has a representation in GitHub, it can be later harvested by specifically designed workflows. For each of the components, a structured *readme file* exists which contains all relevant metadata information about this component. The structure of the *readme file* represents a (project) specific annotation scheme, which is machine readable and which categories can be linked to other existing ontologies. The Ecosystem knowledge graph relied on a new, recent methods for Knowledge Graph Construction (SPARQL Anything) (Asprino et al. 2023) which allowed to extract metadata information for the *Ecosystem* directly from the "README.md" files in Markdown format in Github. A rulebook defined how a repository should be set up, dictates how the readme file to be written, and details the annotation scheme. A validator workflow was written to automatically check the quality of the metadata. Based on this networked GitHub structure, a website[19] was designed to support human navigation through the Ecosystem.

It is this website (for a snapshot see Figure 5) which best illustrates the advantages of such an automatization of documentation. The website enables an overview about the output of the project in terms of its core technological components. Navigation (which can be further tailored) allows visitors to see everything related to a pilot or a work package. At the end of the process, the category *Data* contained 36 components, representing the main important data sources for the project. They are all annotated and can be checked for their release data, license information and so on. The last version of the Data Management Plan (DMP) was in essence built on information from this website (Scharnhorst et al. 2023). But its advantage compared with usual DMP's is that it shows research data in context with other elements in a knowledge production process – research software development (*Tools*) and other forms of documentation (*Reports*). For the intrinsic project management, the Polifonia Ecosystem implementation became the *litmus test* for the readiness and technological maturity of components.

---

[18] personal communication Enrico Daga
[19] https://polifonia-project.github.io/ecosystem/ Access July 12, 2024 Perma link
https://web.archive.org/web/20240712114016/https://polifonia-project.github.io/ecosystem/

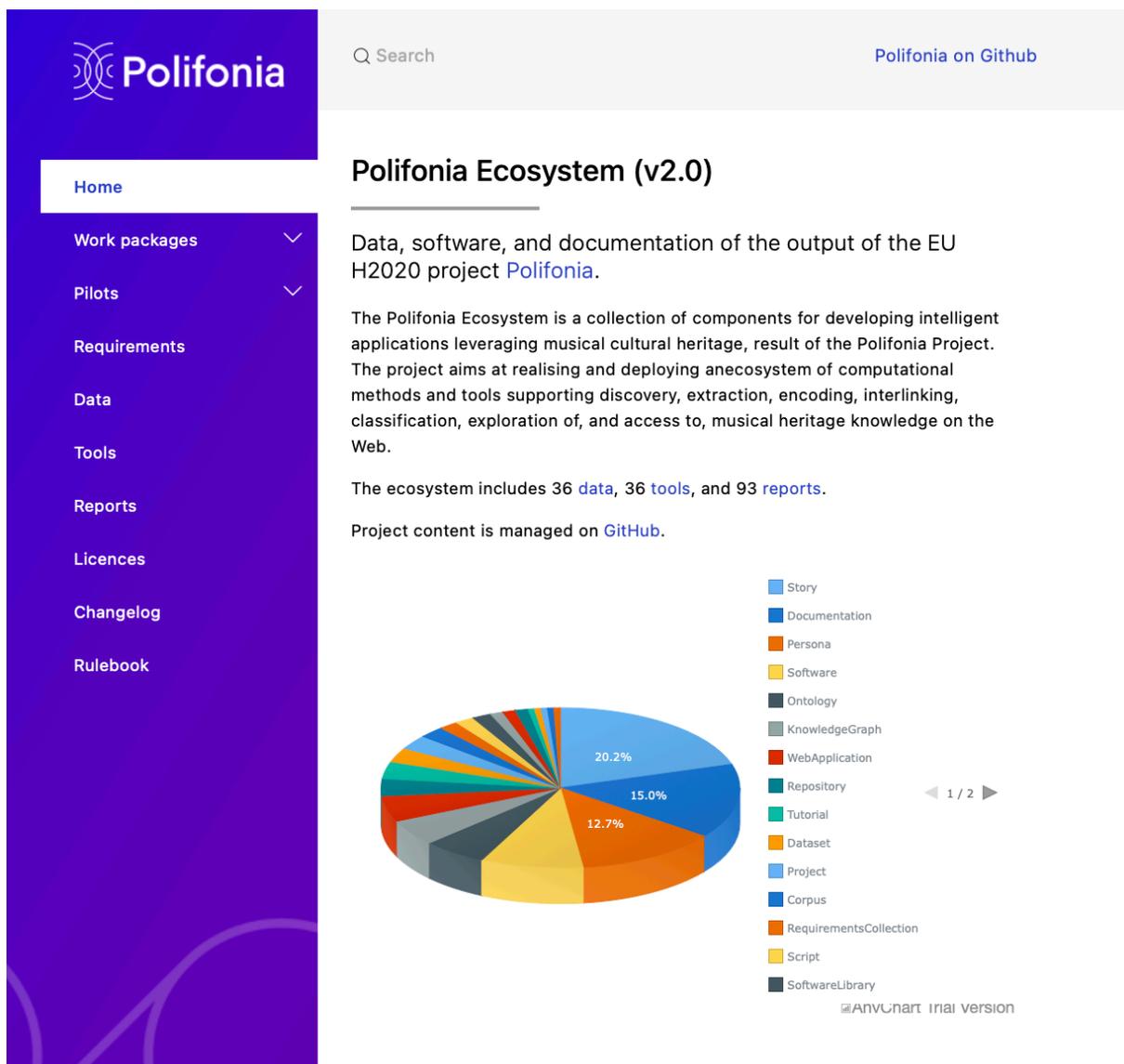

Figure 5: Snapshot of the Polifonia Ecosystem Website

The Polifonia Ecosystem approach is but one example where researchers and information specialists try to develop smart solutions for agile and just-in-time project management. What I hope also to have conveyed in this detailed description is the amount of thinking, planning, designing and adapting going into such an endeavour. I learnt in this project that you need to have a large part of the information gathering process sorted out <u>before</u> you start to turn it into anything computational, machine-readable material. But you also need to be prepared to adapt your design based on experiences with the first implementation steps. An automatic – truly automatic – documentation of the research process will never produce documentation good enough for insightful quantitative studies. The paradigm of Open Science created an extra push for this project to thoroughly formalize its internal project management. However, the documentation would not have reached this level of maturity if not also serving project goals beyond the RDM. Eventually, the Polifonia Ecosystem turned from the task executed by the Technical Board to something the whole Polifonia network embraced, created and advocated.

The lesson learned here: **Good measurement requires well (designed) documentation**. There is always competition between efforts for documentation and efforts for research in research projects. **The benefit of quantifying needs to outweighs the costs of the documentation**. Documentation without design, which may as well just be tracing, will not be good enough for valid insights.

Indeed, you can decently measure something as complex as a research process. But you need to be very clear about the amount of documentation you actually can manage to produce. This is very much in line what Loet and myself formulated as pre-requirement for any new form of studying innovation and other emerging phenomena. Between an idea about a new phenomenon which might be of interest to be further studied and its concrete empirical analysis, there comes the step of operationalization.

> "A crucial step in proceeding from the formulation of analytical hypotheses to the collection of empirical data is implied by the concept of operationalization. How can one move from the analysis to the indication of the importance of the concepts in a social reality? How can a reflexive analyst make a convincing argument when the notion of a system of reference can always be deconstructed, and the time line can be inverted in terms of what the historical account means for the present?
> As systems that contain knowledge should not be considered as given or immediately available for observation, one has to specify them analytically before they can be indicated or measured. In the end, the quantitative analysis depends on the qualitative hypotheses." (Leydesdorff, Scharnhorst 2003, p.13)

## Lessons learned and conclusion

The lessons learned from the three cases discussed above have a certain overlap and contribute to one overall message, even if this message comes in different forms. In the example of the DANS archival service, it appeared that we often have a lot of numbers, but just not the numbers we really need. In the DARIAH case of Key Performance Indicators, we saw that sometimes the categories (variables) you want to measure are more important than the actual numbers. And in the case of the Polifonia project and its Ecosystem approach to Research Data Management, it appeared that in the end the benefits of setting up a system for quantification need to be higher than the costs of the documentation. The core message from all three is thus: without reflection, it is impossible to measure in a meaningful way. When we want to measure, the necessary data are not always available, or even more so, they are almost never available in spite of the many traces we have. What we want to measure might also require far too much time and we can easily find ourselves in a kind of documentation trap. All of these insights are not new. What might be new is our own illusion that this problem could be different now due to digitization and automatization. In the "Measuring the knowledge base" text we stated:

> "Data may also confuse us; particularly when they are so abundantly available as nowadays. A variety of representations is always possible and this problem is further aggravated when databases are no longer substantively codified and dedicated, but when algorithmic search engines and meta-crawlers become widely available. The data provide us first with variation and therefore uncertainty, and the perspectives on the data may also be at variance." (Leydesdorff, Scharnhorst 2003, p. 15)

Indicator research traditionally emerged from established information systems. New information systems related to new emergent structures in research and research infrastructures in turn create possibilities for new indicators. But, they also come with challenges to define those new indicators and to establish effort-efficient ways to measure them.

> "In many cases, one can build on existing definitions of systems—like "the research system" or "the patent system"—but in the case of knowledge-based systems one may also be interested in "emerging systemness". (Leydesdorff, Scharnhorst 2003, p.11)

All three cases (the archive, the ERIC, the research project) are examples of emerging systemness. To understand them clearly, the dimension of time is relevant. One also has to be aware that new phenomena often emerge at interfaces between systems until the point where systems overlap and interact with each other.

> "Emergence can only be analyzed by observing the interaction among systems over time. From this perspective one can analyze the evolution of each system along a trajectory. However, one can also focus on the interaction and potential co-evolution between systems?" (Leydesdorff, Scharnhorst 2003, p. 11)

Since the beginning of using measurements, the makers of indicators (whether new or traditional) always knew **that indicators are pointers** not hard facts. Even more general, all scholars know that **measurements should trigger our thinking and not replace** it. It is, I think, also good to remember ourselves that in academia measurement never stands alone. **It's always serving something**: be it the proof for an existing theory or a catalyst for a new theory. When it comes to empirical studies, they can have different purposes depending of the domain and context in which they are executed. It is important to be explicit about this purpose: what we want to measure in order to understand and what we want to measure in order to manage are two nearly orthogonal dimensions.

When it comes to the connection to between measurement and theory, we also lack theories, or formulated differently, we don't lack them but rather sometimes we have too many of them. I state this not to argue in favor of a *supertheory*. Life is too complex to explained with one theory only, and progress in the sciences lies in our ability to really drill deep in some areas and to build bodies of knowledge which are very specific. But the price we pay for this intimate knowledge is that, on a phenomenological level, things no longer line up and are no longer connected. We miss bridges, translations, connectors between those special bodies of knowledge and thus might become trapped in the arbitrariness of representations.

The 'Measuring the knowledge base' text stated at the beginning of the 2000s:

> "Indicator research so far has reacted sensitively on the changes in knowledge production. New indicators have been proposed regularly. The growing field of webometrics has witnessed an "indicator flood" in an increasingly information rich and knowledge-based environment." ((Leydesdorff, Scharnhorst 2003, p.37)

This holds even more so for the decades to follow publication, which saw and continue to see new data collections and data collecting methods appearing – with generative AI just starting to make its mark on formal scholarly communication and documentation.

Back then, we wrote on the increasing rift between qualitative and quantitative studies, between measurement and theoretical reflection.

> "During the last two decades, the qualitative and the quantitative traditions in science and technology studies have grown increasingly apart (Leydesdorff & Besselaar, 1997; Van den Besselaar, 2000 and 2001). It is time for the pendulum to be turned given the urgent need to understand the effects of different forms of communication and their interaction in knowledge production. In our opinion, the growing diversification and specialization in the sciences, and the relationships to their societal environments, in a knowledge-based economy calls for integrative approaches with detailed appreciation of the ongoing processes of differentiation." (Leydesdorff, Scharnhorst 2003, p. 38)

This remains a persistent problem, and the counter-measure we proposed lies in reflexivity, in the combination of fundamental, theoretically based reflections, methodological explorations and applications in various practices.

> "The cure is discursive reflexivity. Indicator research is not a discipline with a single and commonly accepted theoretical background. It is an "interdiscipline" with approaches as different as the disciplinary background of the researchers. What can still be justified in the case of research results presented to one scientific community of specialists—when the theoretical foundations provide a common basis so that they do not have to be repeated—may loose its justification when crossing a (sub)disciplinary boundary.
> If the theoretical backgrounds in indicator research are not sufficiently reflected, it becomes impossible to create "trading zones" (Nowotny et al., 2001). These trading zones are needed in order to create a dialogue between different approaches inside the branch of quantitative analysis (e.g., between simulation studies and measurement efforts) as well as towards qualitatively oriented science, technology, and innovation studies. New research questions can then be formulated that appreciate the previously achieved results." (Leydesdorff, Scharnhorst 2003, p.38/39)

This call for 'reflexive indicator research' has been taken up inside of the field of quantitative studies of science on a broad front, as the discourses around responsible metrics (Hicks et al. 2015) and new forms of metrics testify (Wilsdon et al. 2017). But the Turin workshop (Petrovich 2024) also demonstrated that there exists a persistent need to nurture a reflexive perspective which, normatively spoken, should be made explicit in any quantitative study around the science system. Naturally, this reflection is rooted in philosophy of science. It enables us to describe the various conceptual reference systems and concrete methods in order to unveil the epistemic foundations of each attempt to measure new phenomena. This is not an easy task.

> "The innovated systems absorb knowledge by being innovated. The observable arrangements therefore have an epistemological status beyond merely providing the analyst with one or another, as yet unreflexive starting point for the narrative..... This research program begins with expectations as different from observations: methodologically controlled observations can then inform the theoretical expectations." (Leydesdorff, Scharnhorst 2003, p. 32)

The three cases I reported represent, in essence, observations. Archives, ERIC's, and research project management are not yet broadly studied phenomena with methodologically controlled observations informed by theoretical expectations. Nevertheless, when writing this paper, I realized how much my own quantitative explorations are still informed by the theoretical debates I had with Loet and other colleagues over decades. It is important for the field of Quantitative Studies of Science to stay in touch with the various practices in the areas supporting research which all make use of some measurements. To join forces with the practitioners, those actors behind the emergence of new institutional phenomenon, can lead to new insights into science dynamics and to better science management. Here, the call of the 'Measuring the knowledge base' text is still valid:

> "What is indicated by the indicator and why is this indicator more suited to that purpose than comparable ones? There is an intrinsic need of validation studies within the indicator domain that is reflexive on the dynamics of the systems that are indicated. Reflexivity gains a particular urgency in phases of de-stabilization, re-organization and the emergence of new (and potentially innovative) structures. When the communication structures are developing at the same time, the starting points or the systems of reference have to be made as clear as possible so that one can trace the changes that are under study in relation to the changes that are made visible and/or explained by the study. This reflexivity can be elaborated in each of the three dimensions: theoretically, historically, and empirically." (Leydesdorff, Scharnhorst 2003, p. 37)

The "Measuring the knowledge base' programme defines the role of communications as the means of self-organisation of social and cognitive processes serious, and reminds us to first raise a more fundamental question:

> "what is communicated?—e.g., economic expectations (in terms of profit and growth), theoretical expectations or perhaps scenarios of what can technologically be realized given institutional and geographic constraints—the focus is firmly set on the specification of the media of communication. How are these communications related and converted into one another? Why are these processes sometimes mutually attractive and reinforcing one another, and under what conditions can the exchanges among them be sustained?" (Leydesdorff, Scharnhorst 2003, p. 37)

Loet himself, over decades, followed the lines set out in this programmatic text. He developed many new methodological approaches to follow the emergence of something new – new communities, new research lines, new ways new ideas turn into innovations, and new ways the sciences organize themselves. He also stood true to the principle to combine theoretical and empirical research. His last book (Leydesdorff 2021) is a testimony of what is possible to unravel by means of sophisticated mathematical tools but never without philosophical reflection. One main requirement posed in the programmatic text 'Measuring the knowledge base' remains:

> "given the drive by the data in indicator research and the development towards an algorithmic understanding, the strengthening of a theoretical approach may become a necessary condition for the further development of quantitative approaches to the study of science, technology, and innovation." (Leydesdorff, Scharnhorst 2003, p. 33)

We need to keep asking 'why do we want to measure'? The text 'Measuring the knowledge base' proposed a new form of innovation studies. While many of its intentions are already realized (EU-SPRI 2024), its call for a comprehensive approach to innovation studies spanning science, economy, politics and any other relevant societal subdynamics (to talk with Luhmann) remains valid.

I miss Loet and our discussions very much. But the beauty in scholarly communication lies in the fact that we can continue to be in discourse with each other even if some actors of a discourse are long gone. Here, Loet was right about the power of communications. I try to comfort myself, and us as community, with fact that the testimony of actors from past and present in form of communications both invites and challenges us to return and inspect previous ways of thinking, respecting the long waves of thoughts documented through human history (Chavalarias et al. 2021). We might profit from excavating those long threads in both our individual- and community-based journeys.

## Acknowledgement
A workshop with the title "Philosophy of Science Meets Quantitative Studies of Science" organized by Eugenio Petrovich in Turin in May 2024 (Petrovich 2024) gave me the possibility to share first ideas about a retrospection of the Leydesdorff/Scharnhorst programme with a wider audience. Furthermore, I'm very grateful for comments given by Kathleen Gregory, Christine Borgman, Sally Wyatt, Enrico Daga, and Paul Wouters on earlier versions of this paper. I would like to thank my colleague Kim Ferguson for her support with copy-editing.